\documentclass[journal=jacsat,manuscript=communication]{achemso}
\usepackage{setspace}
\usepackage[version=3]{mhchem} 
\usepackage[T1]{fontenc}    
\usepackage{color,soul}

\setkeys{acs}{articletitle = true,super = false}
\usepackage{upgreek}
\usepackage{amsmath}
\usepackage{amssymb}
\usepackage{appendix}
\usepackage{esint}
\usepackage{graphicx}
\usepackage{gensymb}
\usepackage{epstopdf}
\usepackage{upgreek}
\usepackage{color}
\usepackage{booktabs}
\usepackage{capt-of}
\usepackage{subfig}

\author{Tao Chen}
\affiliation[GAU]{Third Institute of Physics -- Biophysics, Georg August University, Friedrich-Hund-Platz 1, 37077 G\"{o}ttingen, Germany}
\author{Arindam Ghosh}
\affiliation[GAU]{Third Institute of Physics -- Biophysics, Georg August University, Friedrich-Hund-Platz 1, 37077 G\"{o}ttingen, Germany}
\author{J\"org Enderlein}
\affiliation[GAU]{Third Institute of Physics -- Biophysics, Georg August University, Friedrich-Hund-Platz 1, 37077 G\"{o}ttingen, Germany}
\email{jenderl@gwdg.de}
\alsoaffiliation{Cluster of Excellence ``Multiscale Bioimaging: from Molecular Machines to Networks of Excitable Cells'' (MBExC), University of G\"ottingen, Germany}

\title{Graphene-Induced Energy Transfer for Quantitative Membrane Biophysics at Sub-Nanometer Accuracy}

\date{today}
\begin{document}
\begin{tocentry}
\centering
\includegraphics[width = 0.95\columnwidth]{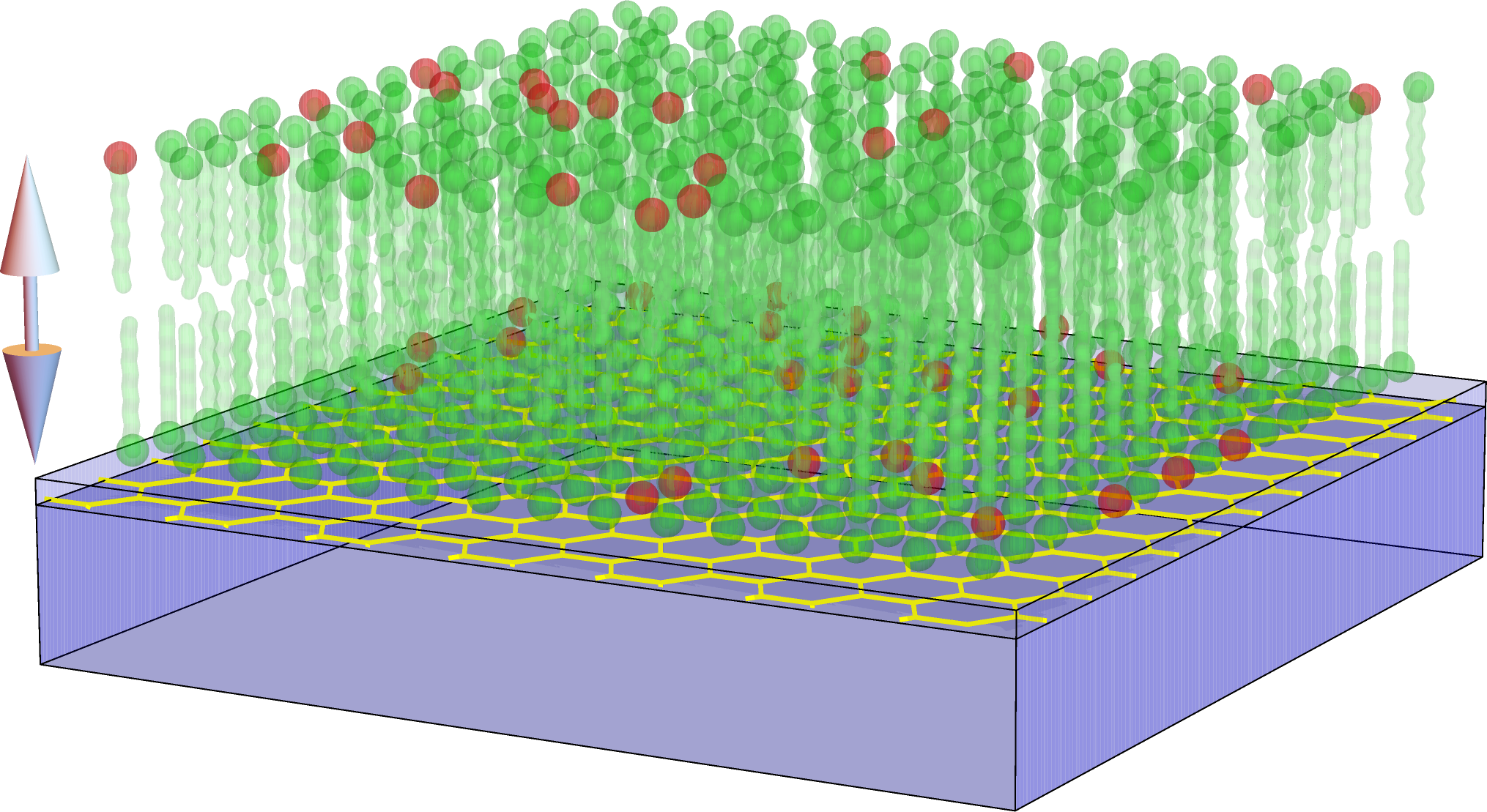}
\end{tocentry}
\begin{abstract}
\noindent
\textit{Graphene-induced energy transfer (GIET) is a recently developed fluorescence-spectroscopic technique that achieves sub-nanometric optical localization of fluorophores along the optical axis of a microscope. GIET is based on the near-field energy transfer from an optically excited fluorescent molecule to a single sheet of graphene. It has been successfully used for estimating inter-leaflet distances of single lipid bilayers, and for investigating the membrane organization of living mitochondria. In this study, we use GIET to measure the cholesterol-induced subtle changes of membrane thickness at the nanoscale. We quantify membrane thickness variations in supported lipid bilayers (SLBs) as a function of lipid composition and increasing cholesterol content. Our findings demonstrate that GIET is an extremely sensitive tool for investigating nanometric structural changes in bio-membranes.}
\end{abstract}

\noindent
\textbf{\large{L}}ipid membranes are one of the fundamental building blocks of cells and form the division between a cell's interior and its environment and between different sub-cellular compartments (mitochondria, nucleus, endoplasmic reticulum etc.). They are composed of lipid bilayers which are decorated with transmembrane and peripheral proteins that are involved in a myriad of different processes and functions (sensing and signal transduction, ion and nutrient transport, endo- and exocytosis etc.). The lipid bilayers themselves are made of amphipathic phospholipid molecules. Among them, cholesterol (Chol) is most abundant ($\sim$30 $\%$) and plays a crucial role in defining a membrane's architecture and dynamics \cite{yeagle1985Chol, maxfield2005role, silvius2003role}. Previous studies have shown that Chol increases lipid ordering in membranes, thus decreasing the area per lipid molecule and increasing the overall thickness of the membrane \cite{hung2007condensing, de2009effect, chakraborty2020Chol, chng2022modulation}, which has direct impact on membrane function \cite{wennberg2012large, armstrong2012effect, sheng2012Chol, kramar2022effect}. For measuring subtle changes in membrane structure and thickness, different experimental methods have been used such as small-angle x-ray scattering (SAXS) \cite{salditt20192} or small-angle neutron scattering (SANS) \cite{gallova2004effect}. These experimental studies are complemented by all-atom and coarse-grained molecular dynamics simulations \cite{de2009effect, berkowitz2009detailed, diaz2015quantifying, wennberg2012large} for providing insight into the structural organization of lipid bilayers with \AA ngstrom resolution. Complementary, fluorescence spectroscopy has been widely used to study membrane dynamics (lipid lateral diffusivity, local membrane viscosity) \cite{eggeling2009direct}, but it is not easy to see with structural variations on the length scale of membrane thickness ($\sim$5~nm). For that purpose, one has to use F\"orster resonance energy transfer (FRET) spectroscopy, \cite{roy2008practical} which requires  labeling with two different fluorescent probes, but even then, measuring membrane thickness values with \AA ngstrom resolution remains challenging. One reason is the difficulty of making FRET measurement of length values quantitatively accurate by taking into account all the complexity of a FRET experiment such as possible direct excitation of acceptor, bleed through of emission from donor, relative sensitivity of detection at different wavelengths, and in particular the usually unknown relative orientation of donor and acceptor molecules with respect to each other and the studied system. 

\begin{figure}[h]
    \centering
    \includegraphics[width = 0.45\textwidth]{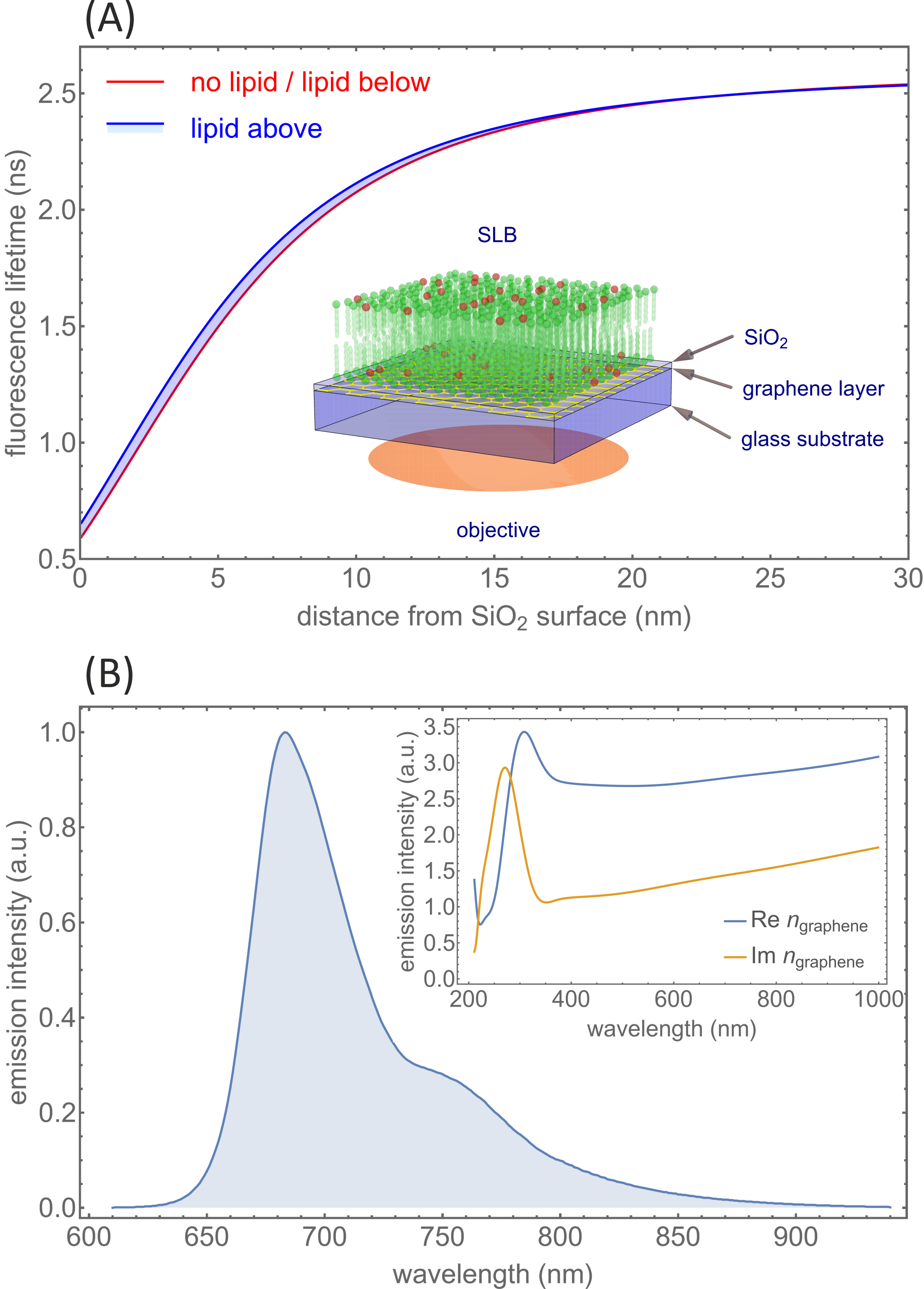}
    \caption{\textbf{Graphene-induced energy transfer (GIET) for membrane biophysics. (A)} GIET calibration curve showing the dependence of the fluorescence lifetime of the dye Atto655 on its distance from the SiO$_2$ surface. To take the effect of a SLB into account, the blue curve shows the result for a $\sim$8~nm thick SLB with refractive index 1.46 \emph{above} the fluorescent molecules (i.e. molecules labeling the head-groups of bottom leaflet lipids). The red curve is the GIET calibration curve for lipids \emph{below} a SLB and is indistinguishable from the calibration curve with no SLB at all. Thus, the blue shaded region contains all curves for dyes \emph{below} a SLB, where the SLB thickness varies from 0~nm (red curve) to 8~nm (blue curve). The inset shows a schematic of the sample/substrate geometry: a single sheet of graphene deposited on top of a glass coverslip and covered with a 10~nm layer of SiO$_2$ layer, on top of which the SLB with few dye-labeled lipids (red dots) is prepared. \textbf{(B)} Emission spectrum of Atto655 and dispersion of the complex-valued refractive index of graphene as used for calculating the GIET calibration curves.} 
    \label{fig:GIET scheme}
\end{figure}

An alternative for measuring distances along one direction with \AA ngstrom resolution is graphene-induced energy transfer (GIET) \cite{ghosh2019graphene, ghosh2021graphene} imaging/spectroscopy. This method is based on the strongly distance-dependent quenching of a fluorescent molecule by a single planar sheet of graphene. This leads to a strong modulation of a fluorophore's intensity and fluorescence lifetime as a function of its distance from the graphene, with full quenching directly on the surface of the graphene and nearly full recovery of bulk fluorescence properties at around 25 nm away from the graphene. Thus, over a range of ca. 25 nm, one can use the graphene-induced quenching for measuring a distance of a fluorophore to the graphene with extreme accuracy by simply measuring its fluorescence lifetime and converting this value into a distance (see supporting info 'Working principle of GIET' and Ref.~\citenum{karedla2014single}). GIET was originally introduced as an advanced variant of the metal-induced energy transfer (MIET) \cite{chizhik2014metal}, where one uses a metal film instead of graphene as the quenching material which covers a dynamic range of ca. 200 nm but with a correspondingly eight-fold reduced spatial resolution. The high spatial resolution of GIET makes it an ideal tool for resolving structural details in small systems such as protein complexes or lipid membranes within a size range of $\sim$5-20~nm. Previously, we have demonstrated the capabilities of GIET by measuring inter-leaflet distances in supported lipid bilayers (SLBs) \cite{ghosh2019graphene} and by mapping the activity-dependent organization of mitochondrial membranes \cite{raja2021mapping}. In the current work, we use GIET for quantifying thickness variations in SLBs as a function of Chol content with sub-nanometer resolution, which is impossible to do with other fluorescence-spectroscopic techniques.  

The core of GIET is the GIET calibration curve which represents the lifetime-versus-distance dependence, see Figure \ref{fig:GIET scheme}. It is calculated in a semi-classical electrodynamics framework by treating a fluorescent dye as an ideal oscillating electric dipole emitter, and then solving Maxwell's equations in the presence of the GIET substrate/sample for such an emitter as the electromagnetic field source. The inset of Figure \ref{fig:GIET scheme}(A) presents a schematic of the GIET substrate with sample: From bottom to top, the GIET substrate consists of a single sheet of graphene deposited on top of a glass coverslip and then covered with a 10~nm thick quartz layer (SiO$_2$) using chemical vapor deposition (see 'Substrate preparation' in Supporting info). SLBs with few fluorescently labeled lipids (Atto655 head group labeling) are prepared using vesicle fusion (see supporting information 'sample preparation'). From the solution of Maxwell's equations, one can then calculate the total emission power of the emitter which is assumed to be proportional to the \emph{radiative} transition rate form the dye's excited state to its ground state. Knowing also the fluorescence lifetime $\tau_0$ and quantum yield $\phi$ of the dye in free space (with no GIET substrate or sample) allows then to compute the final GIET calibration curve - for the details see section 'Working principle of GIET' in Supporting Information. One additional detail that has to be taken into account is that fluorescent dyes have broad emission spectra, and that the dielectric properties of graphene (i.e. its complex-valued refractive index) are wavelength-dependent, which is taken into account by computing the emission rate as a function of wavelength and then averaging the result with the emission spectrum as weight function. Finally, the result depends also on the relative orientation of the dye with respect to the GIET substrate, so that this orientation has also to be known \emph{a priori} and then used in he calculations. For the Atto655 dye used in the current study, we have determined all these parameters in separate measurements and found the values $\tau_0$ = 2.6~ns, $\phi$ = 0.36, and a fluorophore orientation parallel to the membrane surface.\cite{ghosh2019graphene} The emission spectrum of Atto655 and the graphene dispersion curve required for the calculations are shown in Figure \ref{fig:GIET scheme}(B). When calculating the GIET calibration curve, we took also into account the presence of the SLB with refractive index 1.46. This is shown by the blue shaded region in Figure \ref{fig:GIET scheme}(A) that is covered by GIET calibration curves for dyes in the presence of a SLB on top with varying thickness from 0~nm to 8~nm. Remarkably, the presence of a SLB does \emph{not} affect the GIET calibration curve for dyes above the SLB - these curves are indistinguishable from the GIET calibration curve in the absence of a SLB, see red curve in Figure \ref{fig:GIET scheme}(A). The presence of the SLB can shift the GIET calibration curve for dyes below the SLB by up to 0.6 nm to the left with respect to the calibration curve for dyes above the SLB.

\begin{figure}[ht]
    \centering
    \includegraphics[width = 0.45\textwidth]{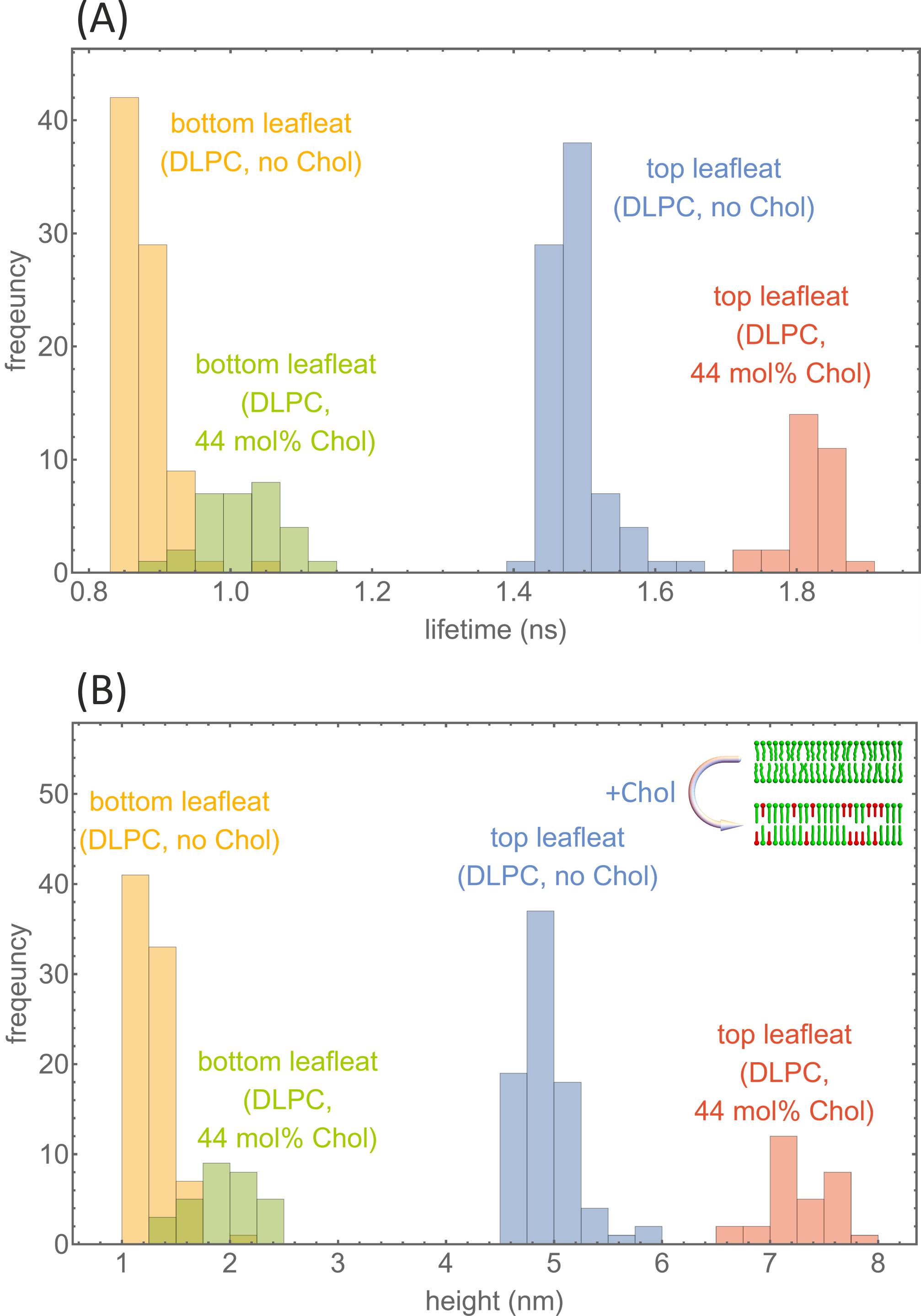}
    \caption{\textbf{GIET measurements on SLBs. (A)} Histograms for determined short and long lifetime values for DOPC SLB without and with 44~mol\% Chol. \textbf{(B)} Histograms of height values derived from the values shown in panel \textbf{(A)} using the GIET calibration curve (for details, see main text). Inset (top right) visualizes that adding Chol changes the lipid organization in an SLB from a \textit{liquid disordered} phase (\textit{L}$_{d}$) towards a \textit{liquid ordered} phase (\textit{L}$_o$).}
    \label{fig:DLPC example}
\end{figure}

For determining values of bilayer thickness, fluorescence decay curves of fluorescently labeled SLBs on GIET substrates were recorded at 20$^\circ$C by scanning areas of 5~$\times$~5~µm$^2$ with a custom-built confocal microscope. We divided each measurement into bunches of 10$^6$ photons and fitted the corresponding TCSPC curves of each bunch with a bi-exponential decay curve to obtain distributions of the short and long fluorescence lifetime component, see an example in Figure \ref{fig:DLPC example}(A) for a DLPC SLB (see below) with no and with 44~mol\% Chol (see Supporting Information section 'Data evaluation' the details of lifetime fitting). These lifetime values were then converted into distance values using the pre-calculated GIET calibration curves. 

The long lifetime components $\tau_\mathrm{long}$, corresponding to the dyes labeling lipid head groups in the top leaflet, were converted into height values $z_\mathrm{top}$ by using the 'no lipid/lipid below' curve shown in Figure \ref{fig:GIET scheme}. However, for converting the short lifetime components $\tau_\mathrm{short}$, corresponding to dyes in the bottom leaflet, into height values $z_\mathrm{bottom}$, we had to use an iterative procedure. In that case, the GIET calibration curve $z_\mathrm{bottom} = g_\mathrm{bottom}(\tau_\mathrm{short},d)$ that relates a measured lifetime $\tau_\mathrm{short}$ to the height value $z_\mathrm{bottom}$ depends on the SLB thickness $d$ itself, which is calculated as the difference between the height value $z_\mathrm{top}$ of dyes above the SLB minus the height value $z_\mathrm{bottom}$ of dyes below the SLB. The height value $z_\mathrm{top}$ is already known from the conversion of the long lifetime component $\tau_\mathrm{long}$. Thus, $z_\mathrm{bottom}$ is found by solving the implicit equation $z_\mathrm{bottom} = g_\mathrm{bottom}(\tau_\mathrm{short}, d)$ with $d=z_\mathrm{top}-z_\mathrm{bottom}$. This is done by iteratively evaluating $z_\mathrm{bottom} = g_\mathrm{bottom}(\tau_\mathrm{short}, d)$ starting with $d=0$ and then updating the value of $d$ with any new value of $z_\mathrm{bottom}$. After 5 iterations, the results for $z_\mathrm{bottom}$ and $d$ do not change anymore. The resulting height histograms for the lifetime data shown in Figure \ref{fig:DLPC example} are presented in Figure \ref{fig:DLPC example}(B).  
    
\begin{figure}
    \centering
    \includegraphics[width = 0.45\textwidth]{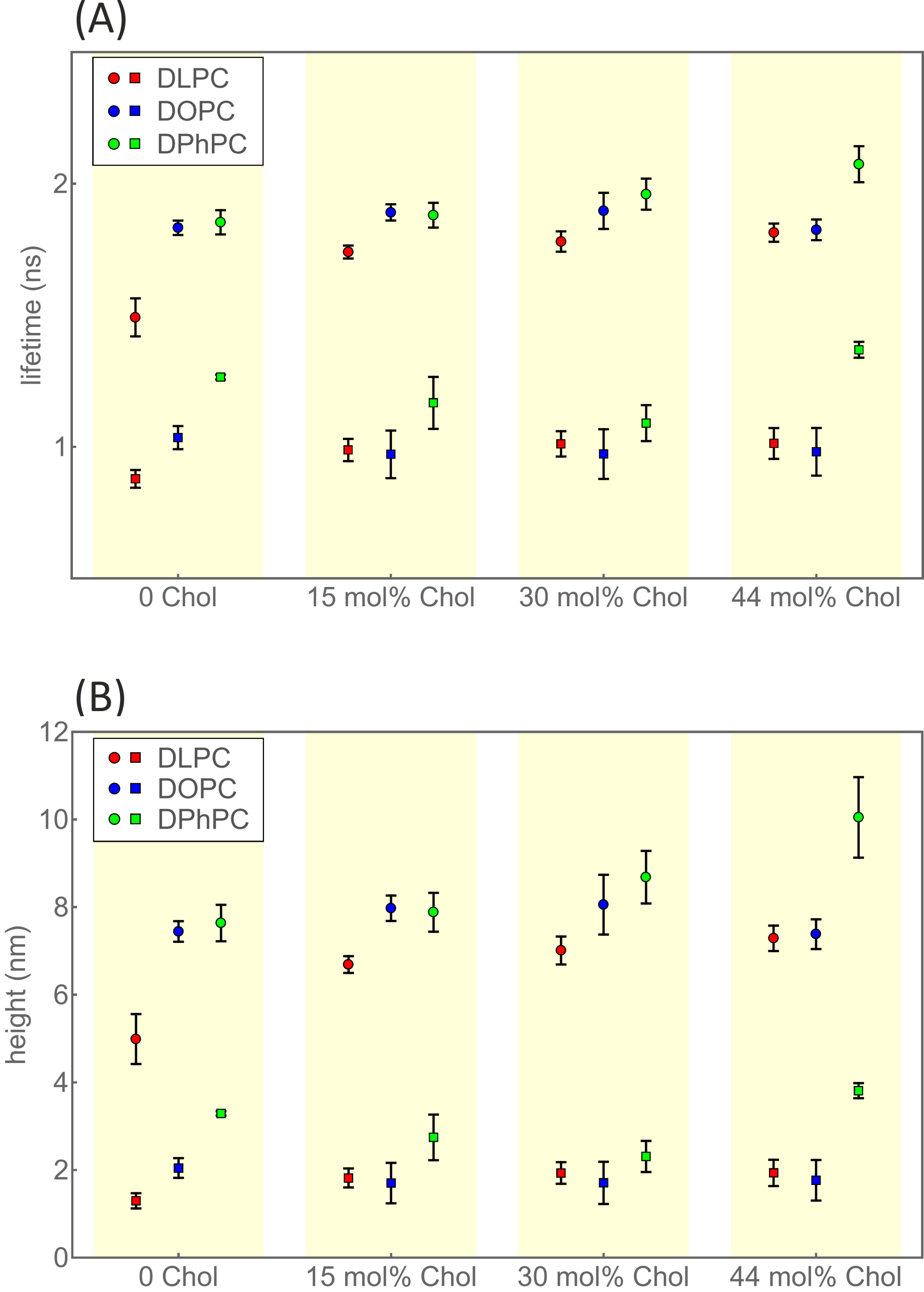}
    \caption{\textbf{Leaflet-specific fluorescence lifetimes and heights as functions of lipid composition.} (\textbf{A}) shows the fluorescence lifetime values obtained from fitting recorded fluorescence decay curves with a bi-exponential fit. The short lifetimes (square markers) refer to dye molecules in the bottom leaflet, the long lifetimes (disk markers) to dye molecules in the top leaflet. Data for the three studied lipid SLBs (DOPC, DLPC, and DPhPC) are grouped according to cholesterol content. (\textbf{B}) Height values derived from the lifetimes of panel (A) by using the GIET calibration curve and adjusted for the lipid bilayer thickness (see main text).}
    \label{fig:Leaflets}
\end{figure}

\begin{table*}
\label{Table 1}
\begin{tabular}{ |p{2.8cm}||p{2.8cm}|p{2.8cm}|p{2.8cm}|p{2.8cm}| }
 \hline
 \multicolumn{5}{|c|}{bilayer thickness (nm)} \\
 \hline
 SLB &No Chol &15 mol\% Chol &30 mol\% Chol &44 mol\% Chol\\
 \hline
 DLPC (12:0)   &3.7 $\pm$ 0.4 (15) &4.9 $\pm$ 0.1 (39) &5.1 $\pm$ 0.4 (18) &5.4 $\pm$ 0.4 (30)\\
 DOPC (18:1)   &5.4 $\pm$ 0.4 (29) &6.3 $\pm$ 0.5 (18) &6.3 $\pm$ 0.7 (17) &5.6 $\pm$ 0.4 (20)\\
 DPhPC (16:0)   &4.3 $\pm$ 0.4 (21) &5.1 $\pm$ 0.3 (16) &6.4 $\pm$ 0.5 (18) &6 $\pm$ 1 (15)\\
 \hline
\end{tabular}
\caption{Thickness values of different SLBs as a function of cholesterol content as measured with GIET. Error bars are standard deviations of $N$ fluorescence lifetime measurements containing 1 million photons each. The number $N$ for each thickness determination is given in brackets after each thickness value. }
\label{thicknesstable}
\end{table*}

For studying the dependence of SLB thickness on Chol concentration, we chose three different lipids for preparing SLBs. The first two SLBs were made with the saturated fatty acids 1,2-dilauroyl-sn-glycero-3-phosphocholine (12:0, DLPC) and 1,2-diphytanoyl-sn-glycero-3-phosphocholine (16:0, DPhPC), and the third SLB was prepared with the mono-unsaturated fatty acid 1,2-dioleoyl-sn-glycero-3-phosphocholine (18:1, DOPC). For all three SLBs, we performed measurements for increasing Chol content of 0, 15, 30, and 44 mol\%. DPhPC consists of saturated branched chain hydrocarbons, and is mainly found in archaebacterial membranes. DPhPC is known to readily form bilayers \cite{lindsey1979physicochemical} and has been used for preparing model membranes in previous studies \cite{veatch2006closed, huang1991lipid}. 

\begin{figure}
    \centering
    \includegraphics[width = 0.44\textwidth]{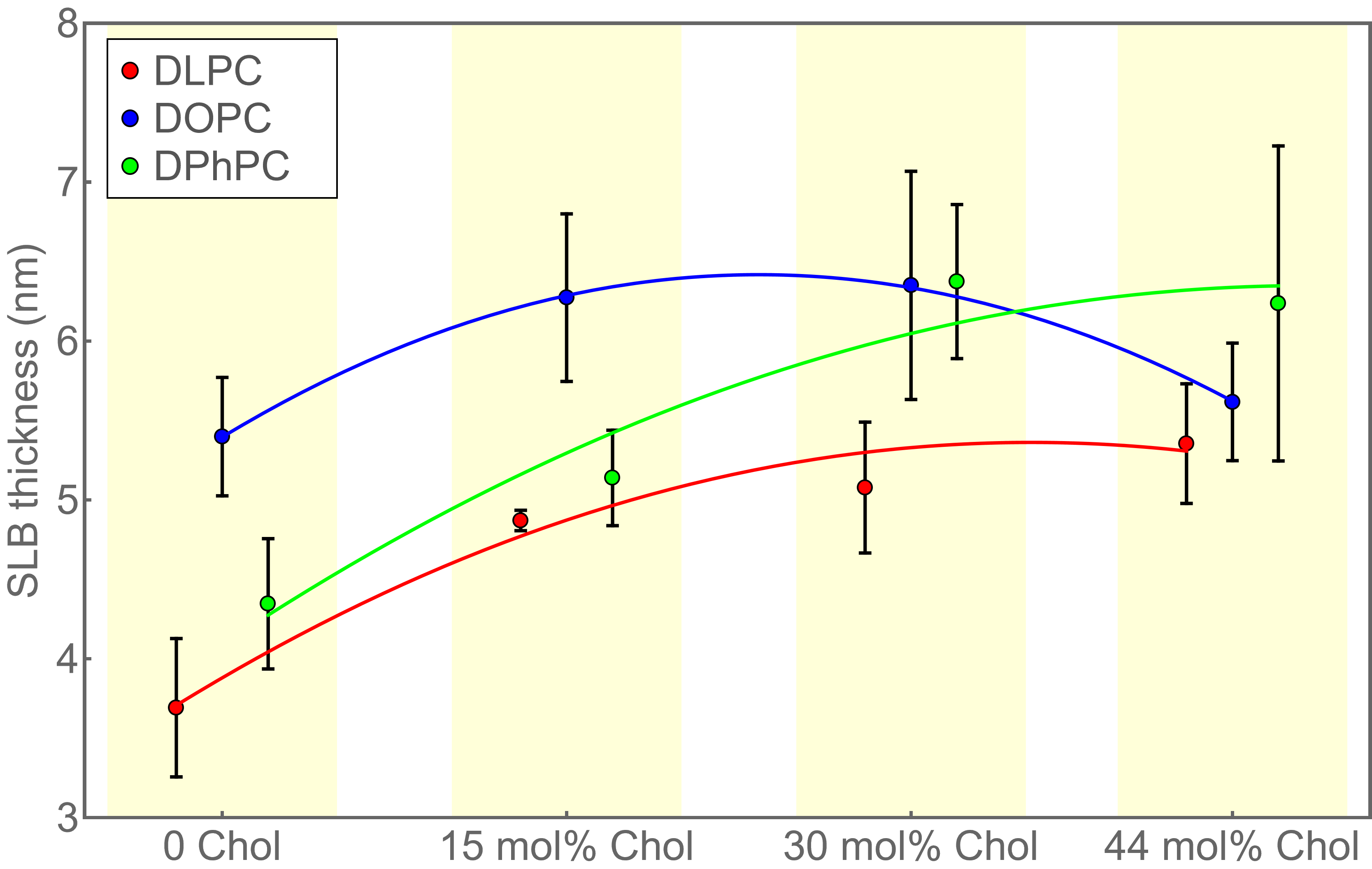}
    \caption{\textbf{Thickness of SLBs.} Average values and standard deviations for thickness values of the SLBs DLPC (12:0) (red), DOPC (18:1) (blue) and DPhPC (12:0) (green) as a function of Chol concentration. For better visibility, data of different SLBs are laterally shifted to each other so that error bars do not overlap. For better visualization, least-square fits of parabolic curves to the data are also shown. These fits shall help to follow the general trend of thickness-vs-Chol concentration dependence, but do not suggest an actual physical square Chol-concentration law of this dependence. Numerical values are listed in Table~\ref{thicknesstable}.}
    \label{fig:Heights}
\end{figure}

We determined the height and resulting thickness values for the twelve SLBs prepared from the three lipids DOPC, DLPC, DPhPC and for four Chol concentrations of 0, 15, 30 and 44 mol\%. The determined height values are displayed in Figure \ref{fig:Leaflets} and listed in the tables in the Supplementary Information. The obtained thickness values are shown in Figure \ref{fig:Heights} and summarized in Table 1. We estimated the errors of our measurements by dividing each measurement into bunches of one million photons, analyzing each bunch separately, and calculating from these results the mean values and standard deviations. For better visualizing the general trend of bilayer thickness as a function of Chol concentration, we fitted these dependencies by parabolic curves, without implying that this is the correct physical model to apply.  

An important result of our GIET-based thickness measurements is that we consistently achieve a measurement accuracy of ca. 0.5 nm (as estimated from the standard deviations). The exception is the DPhPC SLB with 44 mol\% Chol showing a standard deviation of 1 nm. We attribute this to SLB heterogeneity which cannot be spatially resolved by the diffraction-limited lateral resolution of our confocal scanning microscope, but which leads to variations in the determined fluorescence lifetime values. 

Our obtained thickness value for a Chol-free DLPC SLB is in excellent agreement with previous findings from x-ray scattering and independent GIET measurements \cite{kuvcerka2011fluid, ghosh2019graphene}. Moreover, literature values do also suggest a nonlinear increase in bilayer thickness of DLPC SLB with increasing molar fractions of Chol \cite{gallova2004effect}, in accordance to our current findings. It should be mentioned that the length of the linker between dye and lipid head-group as well as the size of the dye itself increase the apparent bilayer thickness as measured by our method, so that our values are systematically larger (and have to be) than those determined by SAXS and SANS, which measure the direct lipid head-group to head-group distance. Literature values of bilayer thickness for Chol-free DPhPC SLBs range from 3.6-4.4~nm \cite{tristram2010structure, lee2005many, wu1995x}, in good agreement with our value of 4.3 $\pm$ 0.4~nm. However, no literature reports for Chol-containing DPhPC SLBs are available. The most interesting behavior show DOPC SLBs, with an increase of bilayer thickness for small Chol concentrations and a decrease at large Chol concentrations. These results are in good agreement with previously reported experimental and computational studies \cite{alwarawrah2010molecular, olsen2013structural}, showing a similar trend in thickness variation of DOPC SLB as a function of Chol content. Furthermore, the thickness values for Chol-free DOPC SLBs agree well with literature values obtained by using atomic force microscopy \cite{attwood2013preparation}, and with our previous measurements using GIET \cite{ghosh2019graphene}.  

Above the phase-transition temperature, SLBs are supposed to exhibit a liquid-disordered ($L_d$) phase, where the hydrocarbon tails of lipids are loosely packed and randomly oriented. Adding Chol to an SLB induces the formation of a liquid-ordered ($L_o$) phase that will grow in size with increasing Chol concentration ('condensing effect') \cite{de2009effect, hung2007condensing}. In the $L_o$ phase, hydrocarbon tails are extended and more closely packed, and Chol condensing leads to a decrease of area per lipid that is considerably more pronounced than one would expect for the case of ideal mixing, where the area per molecule would be a weighted average of Chol and lipid areas.

Another effect of Chol on membrane organization is that it reduces the tilt angle of lipid backbones (with respect to the direction perpendicular to the membrane) and thus increases the inter-leaflet distance between lipid head-groups \cite{levine1971structure, huang2022fundamental}. This explains the Chol-induced increase in thickness of DLPC and DPhPC SLBs, both composed of saturated lipids where adding Chol increases local order and reduces tilting (condensing effect). For SLBs containing unsaturated lipids, a recent study using neutron spin-echo (NSE) spectroscopy, solid-state nuclear magnetic resonance (NMR) spectroscopy, and molecular dynamics (MD) simulations has shown that Chol stiffens the membrane by increasing its lipid packing \cite{chakraborty2020Chol}. Another atomistic MD simulation of DOPC membranes reported an initial increase of membrane thickness with increasing Chol concentration, followed by a decrease for Chol concentrations above 35~mol$\%$ \cite{alwarawrah2010molecular}. The explanation is that in DOPC membranes, acyl-chains are much less ordered than in membranes containing high concentrations of saturated lipids \cite{oh2021effects, pan2008Chol, capponi2016interleaflet, marrink2008Chol}. When adding Chol, this leads first to a reduction of lipid tilt angle for accommodating more space to incorporate Chol \cite{alwarawrah2010molecular}, which is later counteracted by reducing the order of lipid tails at high cholesterol concentrations \cite{olsen2013structural}. This leads first to an increase of bilayer thickness at moderate Chol concentrations, followed by thickness reduction beyond a critical Chol concentration (thinning effect). Our results show that the transition from thickening to thinning occurs between 30 and 44~mol\% Chol concentration. 

To summarize, GIET is a powerful fluorescence-spectroscopic tool for membrane biophysics. We demonstrated this here by studying the impact of Chol on membrane thickness, showing that GIET can easily resolve changes in membrane thickness with an accuracy of ca. 5 \AA ngstrom. An important advantage of GIET is its experimental simplicity, requiring only a conventional fluorescence lifetime imaging microscope (FLIM). Potential further applications of GIET in membrane biophysics can be the precise localization of membrane-associated proteins (or other biomolecules) with respect to a membrane, or the study of membrane dynamics by combining GIET with fluorescence correlation spectroscopy. \\\

\noindent
\textbf{ASSOCIATED CONTENT}

\noindent
\textbf{Supporting Information}
The supporting information is available free of charge. Reagents, Substrate preparation, Sample preparation, Experimental setup, Working principle of GIET imaging, Data evaluation, table S1 enlisting all fluorescence lifetime values and heights corresponding to bottom and top leaflets for DLPC, DOPC and DPhPC. 

\noindent
\textbf{Present addresses}
\textbf{A.G.}: Department of Biotechnology and Biophysics, Biocenter, University of W\"urzburg, Am Hubland, 97074 W\"urzburg, Germany.

\noindent
\textbf{Notes}
The authors declare no competing financial interests.

\noindent
\textbf{Acknowledgments}
AG and JE thank for financial support by the Deutsche Forschungsgemeinschaft (DFG) through project A06 of the SFB~860. JE and TC are grateful to the European Research Council (ERC) for financial support via project “smMIET” (grant agreement no. 884488) under the European Union’s Horizon 2020 research and innovation program. JE is grateful for financial support by the Deutsche Forschungsgemeinschaft (DFG, German Research Foundation) under Germany’s Excellence Strategy - EXC 2067/1- 390729940.

% \bibstyle{achemso}
% \bibliography{references}

\providecommand{\latin}[1]{#1}
\makeatletter
\providecommand{\doi}
  {\begingroup\let\do\@makeother\dospecials
  \catcode`\{=1 \catcode`\}=2 \doi@aux}
\providecommand{\doi@aux}[1]{\endgroup\texttt{#1}}
\makeatother
\providecommand*\mcitethebibliography{\thebibliography}
\csname @ifundefined\endcsname{endmcitethebibliography}
  {\let\endmcitethebibliography\endthebibliography}{}

\end{document}